\documentstyle[prl, aps, amssymb]{revtex}
\begin{document}
\draft
% comment out wideabs for draft
\wideabs{
\title{
$^{63}$Cu NQR Measurement of Stripe Order Parameter 
in La$_{2-x}$Sr$_{x}$CuO$_{4}$}
\author{A. W. Hunt, P. M. Singer, K. R. Thurber, and T. Imai}
\date{\today}
\address{Department of Physics and Center for Materials Science and 
Engineering, M. I. T., Cambridge, MA 02139}
\maketitle
\begin{abstract}
We demonstrate that one can measure the charge-stripe order parameter in the hole-doped CuO$_{2}$ planes 
of La$_{1.875}$Ba$_{0.125}$CuO$_{4}$, La$_{1.48}$Nd$_{0.4}$Sr$_{0.12}$CuO$_{4}$ 
and La$_{1.68}$Eu$_{0.2}$Sr$_{0.12}$CuO$_{4}$ utilizing the wipeout effects of $^{63}$Cu NQR.  
Application of 
the same approach to La$_{2-x}$Sr$_{x}$CuO$_{4}$ reveals the presence of
similar stripe order for the entire underdoped 
superconducting regime $\frac{1}{16}\lesssim x \lesssim\frac{1}{8}$.  
\end{abstract}
%the } correxponsd to wideabs
}
The mechanism of high $T_{c}$ superconductivity has been a major 
controversy throughout the past decade 
\cite{Cape,Science,PW}.  The complexity of the phase diagram for temperature $T$ and 
hole concentration $x$ makes it difficult to 
identify the key leading toward the 
superconducting mechanism.  In 1995, Tranquada et al. 
\cite{tran1,tran2,tran3} demonstrated that La$_{1.6-x}$Nd$_{0.4}$Sr$_{x}$CuO$_{4}$ 
($x\sim\frac{1}{8}$) exhibits charge-stripe order at $T_{charge}=65$ 
K, followed by a spin-stripe order at somewhat lower temperature, 
$T_{spin}=50$ K.   The discovery of the stripe phase \cite{stripe}  
has added a new feature to the already complex phase
diagram.  Initially, some researchers speculated that the stripe 
order was merely a byproduct of the LTO-LTT (low temperature 
orthorhombic-low temperature tetragonal) structural phase transition 
and was extrinsic to the fundamental physics of high 
$T_{c}$ superconductivity.  However, more recently, 
spin-stripe order was observed in La$_{2-x}$Sr$_{x}$CuO$_{4}$ 
($x$ = 0.12 \cite{suzuki,kimura} and $x$ = 0.05 \cite{wakimoto}) 
and La$_{2}$CuO$_{4+\delta}$ \cite{lee}, two 
materials without a LTO-LTT 
structural phase transition.  This suggests that the low energy
(i.e.{\it dynamic}) incommensurate spin excitations previously observed 
for La$_{2-x}$Sr$_{x}$CuO$_{4}$ over the entire 
superconducting regime 0.05 $\lesssim x \lesssim$ 0.25 
\cite{swc,yamada,gabe} may correspond to the dynamic formation of stripes.  Furthermore, 
recent inelastic neutron scattering experiments have established 
similar incommensurability of spin excitations also in underdoped 
YBa$_{2}$Cu$_{3}$O$_{6.6}$ \cite{tran-mook}.  
These developments raise new hope 
that understanding the mechanism 
of stripe instability and its dependence on $T$ and $x$ is the key to 
establishing a unified picture of the hole-doped CuO$_{2}$ planes 
\cite{Cape,Science,PW}.  

Unfortunately, stripe order has proved elusive, escaping 
experimental detection by various techniques.  For example, $^{63}$Cu 
nuclear spin-lattice relaxation rate $^{63}1/T_{1}$ \cite{imai1,allen} measured 
for another striped material with the LTT structure,
 La$_{1.88}$Ba$_{0.12}$CuO$_{4}$ \cite{mu}, exhibits no critical divergence 
at the stripe transition.  The origin of the 
elusiveness is in the {\it glassy} nature of the stripe transition 
\cite{tran3,mu,buchner}.  
In other words, the critical slowing down of spin-stripe fluctuations 
below $T_{charge}$ is more gradual than ordinary 
magnetic phase transitions involving only the spin degrees of 
freedom.  As a consequence, the apparent critical temperature $T_{spin}$ of the spin-stripe 
order is lower for experimental probes with slower frequency scales, 
i.e. $T_{spin}$= 50 K for elastic neutron scattering ($\sim10^{11}$ Hz) 
\cite{tran1,tran2,tran3}, $T_{spin}$= 30 K for ${\mu}$SR ($\sim 
10^{7}$ Hz) \cite{mu}, and $T_{spin}$=10$\sim$30 K for NQR (Nuclear Quadrupole 
Resonance, $\sim10^{6}$ Hz) \cite{Tou,ohsugi,allen}.  

In this Letter, we propose a new experimental approach to investigate 
stripes that takes advantage of extreme sensitivity of $^{63}$Cu NQR to 
local charge distribution.  We will demonstrate that one can accurately measure the 
charge-stripe order parameter in La$_{1.875}$Ba$_{0.125}$CuO$_{4}$ 
and (La,Nd,Eu)$_{1.88}$Sr$_{0.12}$CuO$_{4}$ based on the {\it wipeout 
effects} \cite{white} of $^{63}$Cu NQR.  
The basic idea is quite simple.  
When charge-stripe order sets in at $T_{charge}$, {\it rivers} of hole-rich 
CuO chains separate three-leg CuO ladders without holes, resulting in 
spatial modulation of hole concentration between 0.5 and 0 holes per Cu atom \cite{tran1}.
This would produce as much as ${\sim}$8 MHz \cite{yoshi} 
of instantaneous spatial variation of the $^{63}$Cu NQR frequency 
$^{63}\nu_{Q}$, which is proportional to the electric field gradient 
\cite{slichter} at $^{63}$Cu nuclei.  
This means that $^{63}\nu_{Q}$ is no longer well defined below $T_{charge}$ 
inside the slowly fluctuating striped domains.  
Moreover, charge order turns on low frequency spin fluctuations \cite{tran3}, and 
consequently the $^{63}$Cu
nuclear spin-lattice and spin-spin relaxation rates $^{63}1/T_{1}$ and 
$^{63}1/T_{2}$, respectively, diverge in the striped domains.  
All of these effects should wipeout $^{63}$Cu NQR signals 
within stripe ordered domains even before the growth of {\it 
static} magnetic hyperfine fields changes the NQR frequency completely below 10 K 
\cite{Tou,allen}.  Therefore, we expect that the wipeout fraction ${F}$ of $^{63}$Cu 
NQR signals (i.e.  the fraction of the lost signal intensity) is a good measure of 
stripe order.  In fact, from comparison with elastic 
neutron scattering data in 
La$_{1.48}$Nd$_{0.4}$Sr$_{0.12}$CuO$_{4}$ we demonstrate that the 
wipeout fraction ${F}$ represents nothing but the order parameter for charge-stripe.  
Applying the NQR approach to La$_{2-x}$Sr$_{x}$CuO$_{4}$, 
we demonstrate that a similar stripe 
instability exists over the entire underdoped superconducting regime
 $\frac{1}{16}\lesssim x \lesssim \frac{1}{8}$, but 
disappears right above the critical hole-concentration $x_{c}=\frac{1}{8}$.  

We grew all of the ceramic samples by standard 
solid state reactions \cite{allen}.  The La$_{1.875}$Ba$_{0.125}$CuO$_{4}$ 
sample was synthesized with 99$\%$ $^{63}$Cu isotope enriched CuO.  
We conducted the $^{63,65}$Cu NQR line shape measurements using the
90-180 degree pulse sequence with a fixed separation 
time $t_{90-180}$, typically $t_{90-180}$=10$\sim$12 $\mu$sec.  
The $^{63}$Cu NQR intensity was 
estimated by integrating the Gaussian fit of the NQR line shape, with 
the overall magnitude corrected for the Boltzmann factor and for the spin echo 
decay time $T_{2}$
measured at the NQR peak.  The latter process is essential 
for accurate measurements of the wipeout fraction $F$ because the spin echo 
decay changes dramatically at $T_{charge}$.  The $^{63}$Cu NQR line shape is composed of 
the so-called A-line and B-line with the relative intensity of the latter 
proportional to $x$, in good 
agreement with earlier reports \cite{yoshi}.  
Since high precision measurements conducted for the 
$^{63}$Cu isotope enriched La$_{1.875}$Ba$_{0.125}$CuO$_{4}$ 
showed identical temperature dependence of the NQR intensity 
for the A-line and the B-line 
between 10 K and 300 K, 
we will focus our attention on the behavior of the more intense A-line.
The $^{63}$Cu NQR 
intensity decreases dramatically below $T_{charge}$ 
(i.e. the wipeout fraction $F$ increases from zero to a 
finite value), but the line width of {\it the observable segments} 
of the sample (HWHM = 0.8$\sim$1.8 MHz) showed only a mild 
increase (up to $\sim$35 \%) with decreasing temperature with no 
dramatic change at $T_{charge}$.  

In Fig.1(a), we present the temperature dependence of the wipeout 
fraction $F$ of the A-line in La$_{1.875}$Ba$_{0.125}$CuO$_{4}$, 
La$_{1.48}$Nd$_{0.4}$Sr$_{0.12}$CuO$_{4}$, 
and La$_{1.68}$Eu$_{0.2}$Sr$_{0.12}$CuO$_{4}$.  
Within our experimental 
uncertainties, $F$ is zero between 300 K and 65 K (i.e. we detect 
NQR signals from the entire sample).  However, both $^{63}\nu_{Q}$ and 
$F$ begin to increase abruptly below $T_{charge}$ = 65 K.  
$^{63}\nu_{Q}$ increases $\sim$1 MHz between 65 K and 10 K showing a 
similar temperature dependence as $F$ \cite{allen}, 
and almost the entire NQR signal disappears by 10 K.  We recall that 
below 10 K, the entire $^{63}$Cu resonance intensity reappears as 
Zeeman perturbed NQR (NQR broadened by static magnetic hyperfine field) 
between 20 and 80 MHz \cite{Tou,allen}.  Coincidence of $T_{charge}$ = 65 K for all three 
materials with different LTO-LTT transition temperatures  
\cite{tran1,buchner} ($\sim$65 K for 
La$_{1.875}$Ba$_{0.125}$CuO$_{4}$ and 
La$_{1.48}$Nd$_{0.4}$Sr$_{0.12}$CuO$_{4}$, 
and $\sim$130 K for La$_{1.68}$Eu$_{0.2}$Sr$_{0.12}$CuO$_{4}$) 
indicates that the LTO-LTT structural transition 
is not the primary origin of the observed anomaly at $T_{charge}$.  
In fact, La$_{2-x}$Sr$_{x}$CuO$_{4}$ exhibits qualitatively the 
same temperature dependence of $F$ without the 
LTO-LTT structural phase transition, as presented in Fig. 1(b).  
Accordingly, we must attribute the observed wipeout to electronic 
effects, as discussed above in the third paragraph.  
In Fig.1(a), we compare the wipeout fraction $F$ with the charge- and 
spin-stripe order parameter determined by elastic neutron scattering 
experiments \cite{tran2}.  (We note that by definition the elastic 
neutron scattering intensity represents the square of the order parameter.) 
{\it The identical temperature dependence observed for NQR and neutron data 
for charge ordering allows us to identify the wipeout fraction $F$ 
as the charge-stripe order parameter.}  We note that none of the samples 
with wipeout effects exhibit divergence of $^{139}$La nuclear 
spin-lattice relaxation rate $^{139}1/T_{1}$ at $T_{charge}$ \cite{chou,allen}.  
Instead, $^{139}1/T_{1}$ exhibits gradual upturn below $T_{charge}$, 
showing a hump below 30 K depending on the hole concentration \cite{chou,allen}.  
This is evidence that Cu spin fluctuations 
do not slow down to NQR frequencies immediately at $T_{charge}$, 
in agreement with $\mu$SR \cite{mu}, ESR \cite{buchner}, and earlier NQR line shape 
measurements \cite{Tou,ohsugi}.   
We also found that the temperature dependence of the charge-stripe order 
parameter $F$ fits reasonably well to the weak coupling BCS theory 
with a gap $\Delta=1.75k_{B}T_{charge}=$ 115 $\pm$ 20 K as shown 
in Fig. 1(a).  This is similar to the case of conventional CDW systems 
\cite{gruner}, and suggests a collective nature of charge-stripe order.    

In Fig.2(a), we present representative results of spin echo decay 
observed for La$_{1.875}$Ba$_{0.125}$CuO$_{4}$.  
The results for La$_{1.48}$Nd$_{0.4}$Sr$_{0.12}$CuO$_{4}$ and 
La$_{1.68}$Eu$_{0.2}$Sr$_{0.12}$CuO$_{4}$ were
similar \cite{allen}.  
Spin echo decay exhibits a dramatic crossover at $T_{charge}$ from the convolution of 
Gaussian-Lorentzian \cite{penn}
to Lorentzian with extremely short $T_{2}\sim$ 15 $\mu$sec ($\ll T_{1}$ 
\cite{imai1,allen}), which gradually increases to 10 K. 
The crossover suggests that above $T_{charge}$ both 
the Redfield $T_{1}$ process (Lorentzian) and the indirect nuclear 
spin-spin coupling between {\it like spins}\cite{slichter} (Gaussian) 
contribute to the spin echo decay \cite{penn}, while 
below $T_{charge}$ the spatial modulation of charge and $^{63}\nu_{Q}$ 
reduces the 
number of {\it like spins} and eliminates the indirect coupling 
effects.  The Lorentzian spin echo decay observed below $T_{charge}$ 
suggests that certain magnetic hyperfine fields $H_{hf}$ at the 
Cu sites \cite{isotope} modulate with a
time scale $\tau_{c}$ satisfying $\gamma_{n}H_{hf}\tau_{c}\ll 1$ 
\cite{slichter}, 
where $\gamma_{n}$ is the Cu nuclear gyromagnetic ratio.
A plausible scenario is that charge-stripes 
fluctuate very slowly even below $T_{charge}$, resulting in
modulation of $H_{hf}$ for the observable 
segments of the sample.  We note that sliding motion of CDW is 
known to cause similar motional narrowing in conventional CDW systems 
\cite{cdw}.  The present case is further complicated by the fact that charge order turns on slow 
spin dynamics \cite{tran3} and by the fact that we are limited to 
observing a shrinking domain that has not yet ordered. 
  
We have demonstrated that stripe fluctuations 
wipe out the $^{63}$Cu NQR signal, and the 
wipeout fraction $F$ is a good measure of the charge-stripe order 
parameter.  Motivated by recent observation of spin-stripe order 
in La$_{2}$CuO$_{4+\delta}$ \cite{lee},
La$_{1.88}$Sr$_{0.12}$CuO$_{4}$ \cite{suzuki,kimura}, 
and La$_{1.95}$Sr$_{0.05}$CuO$_{4}$ \cite{wakimoto}, we have searched 
for charge-stripe order in La$_{2-x}$Sr$_{x}$CuO$_{4}$ with $x$=0.15, 
0.135, 0.125, 0.115, 0.09, 0.07, and 0.04.  Utilizing the wipeout 
effects accompanied by crossover to Lorentzian spin echo decay as criteria, we  have 
constructed the phase diagram for charge-stripes.  

Near the optimum doping level with $x=$ 0.15 ($T_{c}=$ 38 K) and 
0.135 ($T_{c}=$ 35 K),  we observed robust Gaussian behavior as shown 
in Fig. 2b, similar to other high 
$T_{c}$ cuprates \cite{penn}.  Slight change in the form of 
non-Lorentzian spin echo decay near $T_{c}$ \cite{Tc} and existence of 
bulk superconductivity make estimation of NQR intensity somewhat tricky, 
but a very safe estimate of the upperbound of 
the wipeout fraction $F$ is $\sim$0.3.  

The sample with $x$ = 0.125 ($T_{c}=$ 31 K) showed pure Lorentzian spin 
echo decay
below $T_{charge}=$ 50 K (not shown, the spin 
echo decay in the striped phase of La$_{2-x}$Sr$_{x}$CuO$_{4}$ is
semi-quantitatively the same as those shown in Fig. 2(a) \cite{allen}).  
This indicates that stripe fluctuations exist in the entire 
sample at 50 K.  
Approximately 40\% of the NQR signal disappears by the superconducting transition 
$T_{c}$.  The sample with 
$x$=0.115 ($T_{c}=$ 31 K) also shows on-set of charge-stripe order at 
$T_{charge}=$ 50 K, and the striped fraction $F$ exhibits a plateau near $T_{c}$, 
where elastic incommensurate magnetic 
scattering also sets in \cite{kimura}.
The behavior of samples with smaller hole concentration, $x$ = 0.09 
($T_{c}=$ 29 K, $T_{charge}$ = 70 K) and 
0.07 ($T_{c}=$ 19 K, $T_{charge}$ = 90 K), 
is qualitatively similar to the case of (La,Nd,Eu)$_{1.88}$Sr$_{0.12}$CuO$_{4}$ 
and La$_{1.875}$Ba$_{0.125}$CuO$_{4}$, but the stripe transition is 
considerably broader.  The broad transition 
may indicate that the stripe is unstable for $x<\frac{1}{8}$. 
Unfortunately, we cannot
determine the real space structure of the stripe \cite{stripe} based on
wipeout effects alone.  
The Meissner fraction in $x=$ 0.125, 0.115, 0.09, and 0.07 \cite{allen} 
is comparable to the observable NQR fraction, $1-F$, at $T_{c}$ for each sample.  
Whether superconductivity coexists with 
stripe, or microscopic phase separation sets in at $T_{c}$, is a matter 
of debate.  However, it is certainly true that a canonical
superconducting transition with a large specific heat jump at $T_{c}$ is realized 
only above $x$ = 0.125 \cite{loram}.

In contrast to the superconducting samples, the lightly hole-doped 
non-superconducting La$_{1.96}$Sr$_{0.04}$ CuO$_{4}$ showed a gradual decrease of 
the NQR intensity below 300 K.  
However, the spin echo decay always has a Gaussian component as shown in 
Fig.2(b), and there is no evidence for stripe 
fluctuations.  This is consistent with the neutron scattering result 
that spin response is commensurate below $x$ = 0.05 without the
 signature of spin stripes \cite{keimer,wakimoto,yamada}.  
Since dc resistivity for $x$ = 0.04 is large and 
shows a gradual upturn due to weak localization 
below 300 K \cite{keimer}, we attribute the 
gradual loss of the NQR signal to commensurate
short-range spin order in the vicinity of {\it randomly} localized holes.
In this context, it is worth recalling that even conventional dilute Kondo alloys such 
as Cu:Fe exhibit similar gradual wipeout ($F\sim\frac{1}{T}$)  
above the Kondo temperature \cite{Tk}.
 
In Fig. 3 we summarize the critical temperature $T_{charge}$ and order 
parameter $F$ as a function of $x$.  Clearly, the ground state of La$_{2-x}$Sr$_{x}$CuO$_{4}$ 
exhibits a sharp crossover at the critical hole 
concentration $x_{c}=\frac{1}{8}$ from the striped state to the superconducting state
with low energy ($\sim$meV) stripe fluctuations \cite{swc,yamada,gabe}.  
The charge-ordering temperature $T_{charge}$ is consistently higher than the spin ordering temperature 
determined by elastic neutron scattering 
($T_{spin}$ = 30 K for $x$ = 0.12 \cite{kimura}, and 17 K for 
$x$ = 0.05 \cite{wakimoto}), 
$\mu$SR and $^{139}$La NQR ($T_{spin}\sim$ 15 K 
is the largest for $x$ = 0.115 \cite{mu,chou,ohsugi,allen}).  
Perhaps remnant fluctuations of charge-stripes prevent spins from freezing, resulting in the glassy 
transition.  The fact that $T_{charge}$ is much higher than $T_{spin}$ 
in La$_{2-x}$Sr$_{x}$CuO$_{4}$ may indicate that the charge-stripes 
are significantly disordered.  The previously 
undetected phase boundary of charge-stripe in Fig.3(b) explains why insulating 
(i.e. divergent) behavior of $^{63}1/T_{1}$ 
\cite{imai} and dc-resistivity \cite{Bo} crosses over 
to metallic in the vicinity of the optimal superconducting 
phase $x$=0.15.  These observations as well as recent neutron data 
\cite{tran1,tran2,tran3,suzuki,kimura,wakimoto,lee,swc,yamada,gabe} are 
compatible with, but not necessarily a proof of, a viewpoint \cite{PW,1/8} 
that stripe fluctuations near $x_{c}=\frac{1}{8}$ have a connection with the mechanism of high 
$T_{c}$ superconductivity.  Even if the stripe instability is actually 
disfavorable for superconducting pairing, the proximity between the 
quasi-static stripe phase and the canonical superconducting phase in Fig.3(b) indicates 
that they are energetically very close.  In any case, the implications of all earlier experiments 
reported for underdoped, superconducting La$_{2-x}$Sr$_{x}$CuO$_{4}$ 
($\frac{1}{16}\lesssim x \lesssim \frac{1}{8}$) during the past decade 
must be reassessed in light of our finding that there exist slowly 
fluctuating, quasi-static charge-stripes.  

We thank Y.S. Lee, M.A. Kastner, R.J. Birgeneau, P.A. Lee, S. 
Sachdev, M. Takigawa, K. Sakaie,
and F.C. Chou for discussions, and B. Batlogg for calling 
our attention to Ref.\cite{loram}.  T.I. acknowledges that his detailed 
knowledge of 214 compounds arises from earlier fruitful 
collaboration with K. Yoshimura and C.P. Slichter.  
This work was supported by NSF DMR 96-23858, NSF DMR 98-08941,
 the A.P. Sloan foundation, the Mitsui foundation, and the Platzman fund.

% now the references. delete or change fake bibitem. delete next three
%   lines and directly read in your .bbl file if you use bibtex.
 
% \begin{figure}
% \caption{}
% \label{}
% \end{figure}

\begin{figure}
\caption{
(a) The wipeout fraction $F$ of $^{63}$Cu NQR 
 in La$_{1.875}$Ba$_{0.125}$CuO$_{4}$ ($\blacktriangle$),
 La$_{1.48}$Nd$_{0.4}$Sr$_{0.12}$CuO$_{4}$ ($\bullet$),  
and La$_{1.68}$Eu$_{0.2}$Sr$_{0.12}$CuO$_{4}$ ($\times$).  The solid curve 
represents the fit to the BCS weak coupling theory with $\Delta$ = 115 K.  
The spin- ($\circ$) and charge-stripe ($\triangle$) order parameter 
in La$_{1.48}$Nd$_{0.4}$Sr$_{0.12}$CuO$_{4}$ %\cite{tran2} 
[3] are also plotted.  
(b) The wipeout fraction $F$ in 
La$_{2-x}$Sr$_{x}$CuO$_{4}$ for x= 0.125 ($\blacktriangle$), 0.115 
($\times$), 0.09 ($\bullet$), and 0.07 ($\blacksquare$).  The curves 
are guides to the eyes.
}
\label{1}
\end{figure}
 
 \begin{figure}
 \caption{Spin echo decay normalized by the Boltzmann factor.  
 (a) La$_{1.875}$Ba$_{0.125}$CuO$_{4}$, and 
 (b) La$_{1.85}$Sr$_{0.15}$CuO$_{4}$ ($\bullet$,$\blacksquare$) and 
 La$_{1.96}$Sr$_{0.04}$CuO$_{4}$ ($\triangledown$,$\Diamond$).  
 Solid curves are the best fits 
 to a convolution of Gaussian-Lorentzian (curved, both the Gaussian and 
 Lorentzian time constants are taken as free parameters) or a pure Lorentzian 
 (straight lines).  Notice that the NQR intensity, $m(2t
 _{90-180}$$=$$0)$ decreases 
 drastically below $T_{charge}$=65K in La$_{1.875}$Ba$_{0.125}$CuO$_{4}$.  
 }
 \label{2}
 \end{figure}

 \begin{figure}
 \caption{(a) The charge-stripe order parameter $F$ at 10 K in La$_{2-x}$Sr$_{x}$CuO$_{4}$.
 (b) The phase diagram of the charge-stripe order $T_{charge}$  
 ($\bullet$,$\circ$) and the 
 superconducting critical temperature $T_{c}$ ($\blacksquare$,$\square$) in 
 La$_{2-x}$Sr$_{x}$CuO$_{4}$.  Filled (open) symbols denote the phase boundary 
 with more (less) than 50\% fraction. 
 }
 \label{3}
 \end{figure}
% \begin{table}
% \caption{}
% \label{}
% \end{tabular}
% \end{table}


\begin{references}
\bibitem{Tc}We saw a hint of change in spin echo decay at $T \gtrsim T_{c}$, which 
might be an indication of stripe fluctuations\cite{allen}.
\end{references}
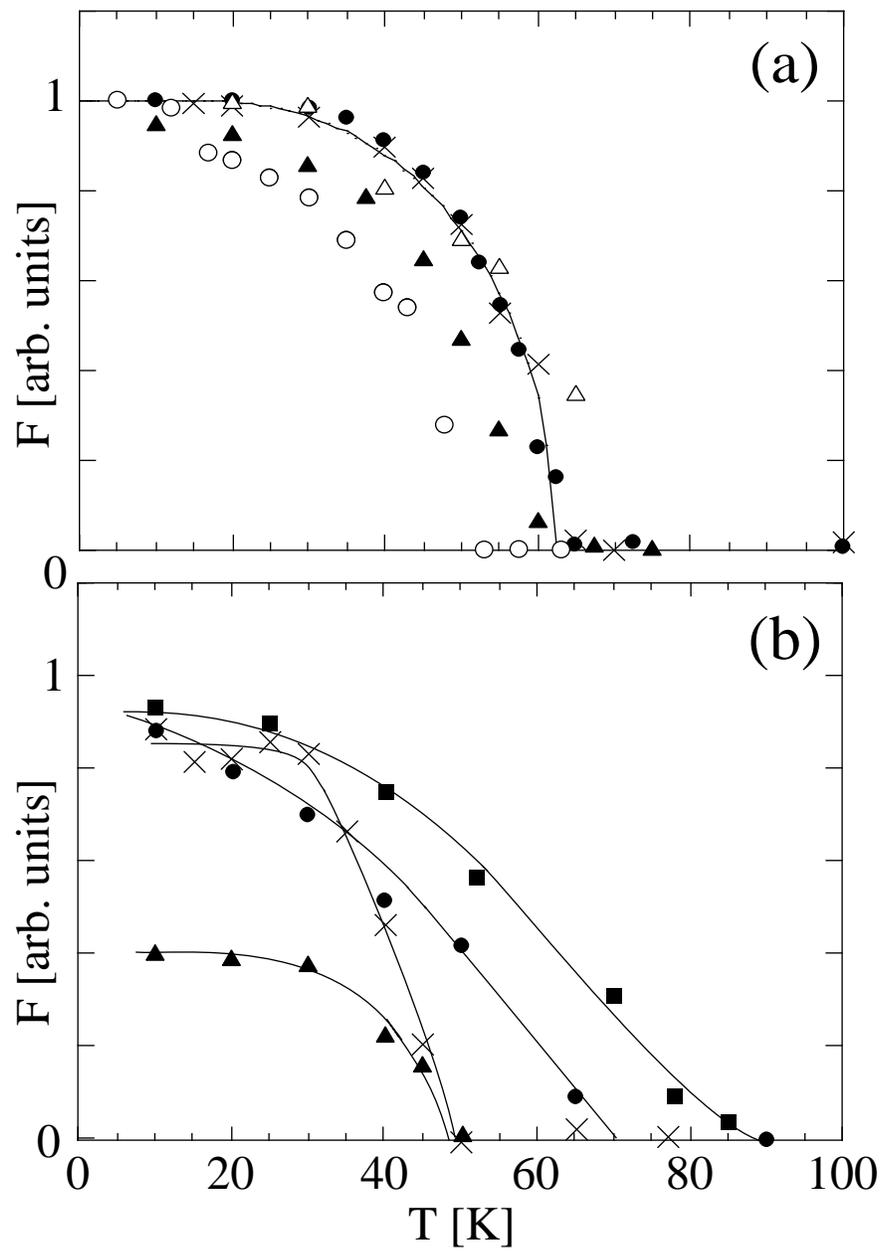
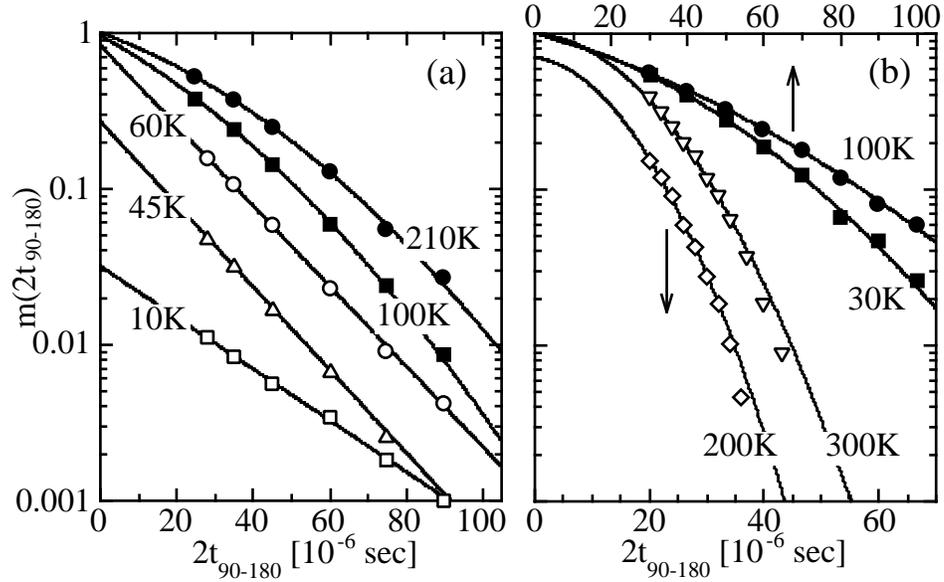
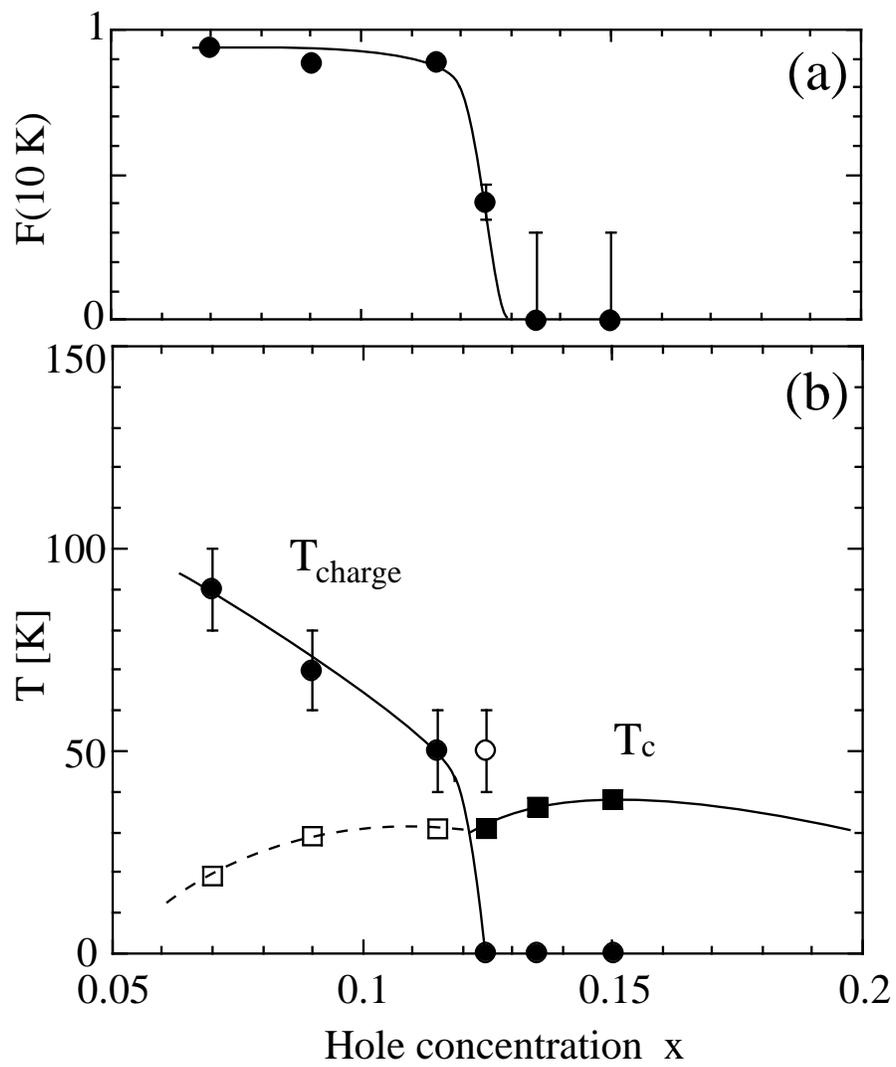
\end{document}